# MAGNETOFLUIDIC BASED CONTROLLED DROPLET BREAKUP: EFFECT OF NON-UNIFORM FORCE FIELD

Sudip Shyam, Bhavesh Dhapola, Pranab Kumar Mondal*

Microfluidics and Microscale Transport Processes Laboratory,
Department of Mechanical Engineering,
Indian Institute of Technology Guwahati, Guwahati 781039, India

**ABSTRACT**

We report the breakup dynamics of a magnetically active droplet (ferrofluid droplet) in a T-shaped LOC device under the modulation of a non-uniform magnetic field. We adhere to high-speed imaging modalities for the experimental quantification of the droplet splitting phenomena in the presence of a non-uniform force field gradient, while the underlying phenomena is supported by the numerical results in a qualitative manner as well. On reaching the T-junction divergence, the droplet engulfs the intersection fully and eventually deforms into the dumbbell-shaped form making its bulges to move towards the branches of the junction. We observe that the asymmetric distribution of the magnetic force lines, acting over the T-junction divergence, induces an accelerating motion to the left moving bulge (since the magnet is placed adjacent to the left branch). We show that the non-uniform force field gradient allows the formation of a *hump-like* structure inside the left moving bulge which triggers the onset of augmented convection in its flow field. We reveal that this augmented internal convection developed in the left moving volume/bulge, on getting coupled with the various involved time scales of the flow field, lead to the asymmetric splitting of the droplet into two sister droplets. Our analysis establishes that at the critical strength of the applied forcing, as realized by the critical magnetic Bond number, the flow time scale becomes minimum at the left branch of the channel, leading to the formation of larger sized sister droplet therein. Inferences of the present analysis, which focuses on the simple, wireless, robust and low-cost droplet splitting mechanism, will provide a potential solution for rapid droplet breakup, typically finds significant importance in point-of-care diagnostics.

**Keywords:** *Breakup, Drops, Micro-/Nano-fluid dynamics, Microfluidics*

*Email address for correspondence: mail2pranab@gmail.com; pranabm@iitg.ac.in*

1. **INTRODUCTION**

Droplet-based microfluidics has gained widespread attention among researchers for the past two decades, owing to its capability of precise control of minute volumes (Adamson *et al.* 2006; Baroud *et al.* 2010; Manga 1996; Vladisavljević *et al.* 2013). Important to mention here that the paradigm of droplet-based microfluidics has shown significant technical and research potential specifically due to its suitability in the point of care diagnostics (Shamloo & Hassani-Gangaraj 2020), drug delivery (Yue *et al.* 2020), analyzing and screening of bio/chemical reaction products (Zheng & Ismagilov 2005) and many more related areas (Madadelahi *et al.* 2019; Moon *et al.* 2010; Santos *et al.* 2016). Researchers have explored several aspects of droplet behavior such as droplet generation, breakup (Hoang *et al.* 2013; Jullien *et al.* 2009; Leshansky *et al.* 2012; Leshansky & Pismen 2009; Link *et al.* 2004), and merging/coalescence (Christopher *et al.* 2009), in various microfluidic devices such as T-junction, flow-focusing junction, Co-flowing junctions, to name a few. Note that the breakup of a mother droplet into two sister droplets in a simple passive microfluidic device such as a T junction, has significant engineering implications. Researchers have shown that the size of the sister droplet being splitted in a T- junction can be passively controlled by adjusting the length of the downstream channel (Link *et al.* 2004). However, situations in which there exist geometric limitations, the suitability of passive breakup methodology of droplets is highly restrained. This limitation, in particular, has led to the paradigm of active droplet breakup mechanism, whereby external field modulated forcing such as electric (Xi *et al.* 2016), magnetic (Tan & Nguyen 2011), acoustic (Schmid & Franke 2013), temperature (Yesiloz *et al.* 2017), optics (Marchand *et al.* 2012) are used for maneuvering the fluid flow field.

It may be mentioned here that in comparison to other fields, utilization of magnetic field for controlling the droplet breakup phenomena has some serious advantageous features. The magnetic field does not induce any changes in the flow field such as pH, ionic concentration, and surface charge. The magnetofluidic based droplet manipulation is usually realized with help of a smart fluid known as ferrofluid. Ferrofluid is a colloidal suspension of ferro/ferri magnetic particles in a non-magnetic carrier medium (Odenbach 2002; Rosensweig 1984). It is worth adding here that ferrofluid exhibits superparamagnetic nature i.e., on application of magnetic field, its magnetization is comparable to any ferromagnetic particles. Whereas on removal of the applied field, the ferrofluid flow field does not display any net hysteresis (Rosensweig 1984). Ferrofluid has been successfully used in many engineering applications such as separation (Hejazian *et al.* 2015), heat transfer augmentation (Shyam *et al.* 2019), mixing (Kitenbergs *et al.* 2015; Zhu & Nguyen 2012), droplet generation (Tan *et al.* 2010), breakup (Bijarchi *et al.* 2021) and many more. Although breakup of droplet in a T-junction has been widely explored, studies pertaining to ferrofluid droplet break up under the modulation of magnetic field is sparsely explored. Having a closer scrutiny of the available literature, it is found that researchers have investigated the implication of uniform magnetic field on the ferrofluid droplet splitting phenomena (Li *et al.* 2016; Ma *et al.* 2017; Wu *et al.* 2013, 2014). It has been shown both numerically and experimentally that under the modulation of uniform magnetic field, the splitting phenomena is mostly symmetric. Also from the reported analysis in this paradigm, it is apparent that the application of uniform magnetic field brings in sufficient control on the size of the sister droplets generated from the mother droplet.

Albeit several underlying issues of the implication of uniform magnetic field on the splitting of the ferrofluid droplet is well explored (sec Refs (Li *et al.* 2016; Ma *et al.* 2017; Wu *et al.* 2013a, 2013b, 2014), research endeavor with an emphasis on exploring the implication of the non-uniformity of the magnetic field distribution on the overall droplet breakup phenomena has been sparsely investigated. It may be mentioned here that uniform magnetic field is a mathematical assumption, while in a realistic physical scenario, a certain amount of non-uniformity is bound to exist pertaining to magnetic field driven fluidic applications. A few researchers, however, have explored the role of non-uniformity of the applied magnetic field on the overall size of the sister droplets generated in the process and discussed the asymmetric splitting phenomena attributing to the onset of the involved forcing gradient (Aboutalebi *et al.* 2018; Amiri Roodan *et al.* 2020; Bijarchi *et al.* 2021).

It is worth mentioning here that literatures exploring the effect of inequality of the force field gradient on the underlying breakup dynamics in the purview of droplet train flow has not been well explored. Due to the non-uniform distribution of magnetic flux density, the sister droplets moving downstream of the left/right branch of the T-junction after the breakup phenomena, will possess different flow time scale. The imbalance of this flow time scale in both the branches is expected to affect the overall droplet splitting phenomena and may result in generation of sister droplet of unequal sizes. The non-uniform force field alterations in droplet splitting/break-up event is expected to be more fascinating in the presence of droplet train flow, attributed primarily to the imbalance between the involved flow time scales. This aspect of droplet splitting phenomena albeit interesting from a fluid dynamics point of view and would be of huge practical relevance in different applications, has not been studied in literature till date.

In the present investigation, we present a novel way of controlling the droplet break up phenomena in a T-junction divergence of LOC device, under the modulation of a non-uniform magnetic field. Droplets are generated in a T-junction of the LOC device, following which they are splitted in another T-junction divergence, located further downstream. A magnet is placed adjacent to the left branch of the T-junction divergence, thereby inducing asymmetric magnetic flux distribution. We demonstrate that the asymmetric force field distribution creates uneven flow time scale in the left and right branches of the divergence (precisely T-junction divergence). We show that by specifically tuning the balance between the various time scales acting on the droplet flow field, we could control the size of the sister droplet. Also, we numerically simulate the flow dynamics under the influence of a magnetic field, essentially for a qualitative understanding of the droplet break-up characteristics in the asymmetric force field ambience. In what follows, we divide this study into four sections. In the first section, we explore the droplet break-up phenomena in the presence of a non-uniform magnetic field. In this section, we discuss the droplet splitting phenomena under the modulation of external force field obtained from both experimental investigations and numerical simulations. In the second section, we numerically explore internal hydrodynamics of the droplet during its splitting under the modulation of a non-uniform magnetic forcing. In the subsequent section, we experimentally investigate the morphological evolution of the droplet splitting characteristics. In the final section of this article, we attempt to develop a physical understanding of the typical droplet break-up behavior by exploiting the various involved time scales.

## 2. MATERIALS AND METHODS
### 2.1 EXPERIMENTAL METHODS

In the present study, we use ferrofluid solution as the dispersed phase, while silicon oil is used as the continuous phase. We employ co-precipitation method for the preparation of ferrofluid solution. The prepared dispersed phase, i.e., the ferrofluid solution is composed of DI-water (De-ionized water) as the carrier phase, while iron-oxide nanoparticles form the suspended phase. Interested readers are referred to our recent articles where the preparation of the ferrofluid solution has been aptly discussed (Shyam *et al.* 2020b, 2020a, 2020c, 2021). Figure 1 depicts the characterization of the prepared ferrofluid sample. The ferrofluid solution exhibits superparamagnetic characteristics as can be observed from the $M-H$ curve of Figure 1(a). We show in Figure 1(b)-(c), the variation of Zeta potential and the size of the suspended nanoparticle in the ferrofluid solution. Note that the ferrofluid solution has a zeta potential of around $-53\ mV$, signifying an electrostatically stable solution (Xu 2002).

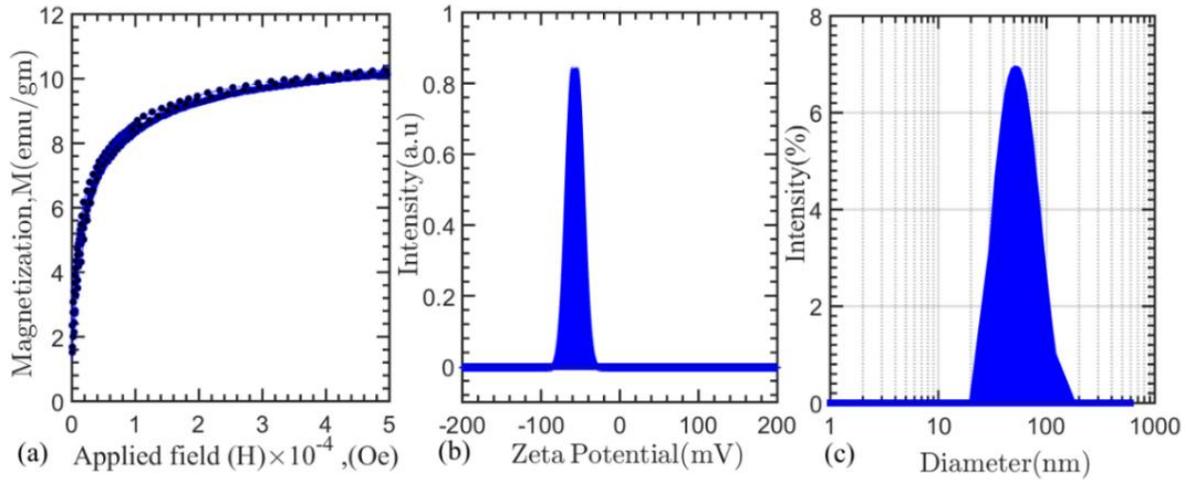

**FIGURE** 1. **(Color online)** Plot depicts the magnetization curve of the prepared ferrofluid sample as was measured by VSM (vibrating sample magnetometer). The prepared ferrofluid sample exhibits no hysteresis and is superparamagnetic in nature, (b) Plot shows the electrostatic potential characteristics of the ferrofluid solution, (c) Plot illustrates the size distribution of the suspended magnetic nanoparticles in the ferrofluid sample. The average size of the magnetic nanoparticles was found to be 50 $nm$.

The density and viscosity of the ferrofluid solution are calculated to be around $1050\ kg \cdot m^{-3}$, and $0.00106\ Pa \cdot s$ respectively. While the volume fraction of the iron-nanoparticles in the ferrofluid solution was around 2 %. As already mentioned, we use silicon oil (make: Sigma-Aldrich) as the continuous phase in the present study. Accordingly, the density and viscosity of silicon oil are found to be around $930\ kg \cdot m^{-3}$, and $0.3\ Pa \cdot s$. The interfacial tension between silicon oil and ferrofluid, as measured by tensiometer (Make: Kyowa), is found to be around $0.012\ N \cdot m^{-1}$. Note that the iron-oxide nanoparticles particles are coated with surfactant (lauric acid for the present case) essentially to avoid any agglomeration, which is not unlikely to occur due to the interparticle interactions (Odenbach 2002; Rosensweig 1984; Shyam *et al.* 2019). The presence of surfactant in the ferrofluid solution lowers its static contact angle on a rigid substrate, as can be clearly observed from Figure 2(a).

The LOC-based device, consisting of a T-shaped fluidic channel, is fabricated by using soft lithographic technique (Whitesides & Stroock 2001). The schematic representation of the microfluidic device is shown in Figure 2(b). The microfluidic passage has a square cross-section of around $100 \, \mu m$ as width. The fabricated device has three section: a droplet generation junction (T-junction), a straight microchannel, and a T junction divergence (in which droplet splitting is taking place). The continuous phase (silicon oil) and the dispersed phase (ferrofluid) are injected from the two inlets leading to the droplet formation at the T-junction (i.e. droplet generation junction). The generated droplet, henceforth will be referred to as the mother droplet, then flows through the straight microchannel and further breaks down into smaller droplets (henceforth, these droplets will be referred to as sister droplets) at the T-junction divergence as can be seen from Figure 2(b). A Neodymium Iron boron magnet (NdFeB) is placed adjacent to the left branch (of the T-junction divergence) for the induction of the non-uniform magnetic field. The presence of the uneven magnetic field gradient ensures asymmetric droplet splitting, as can be observed from Figure 2(b).

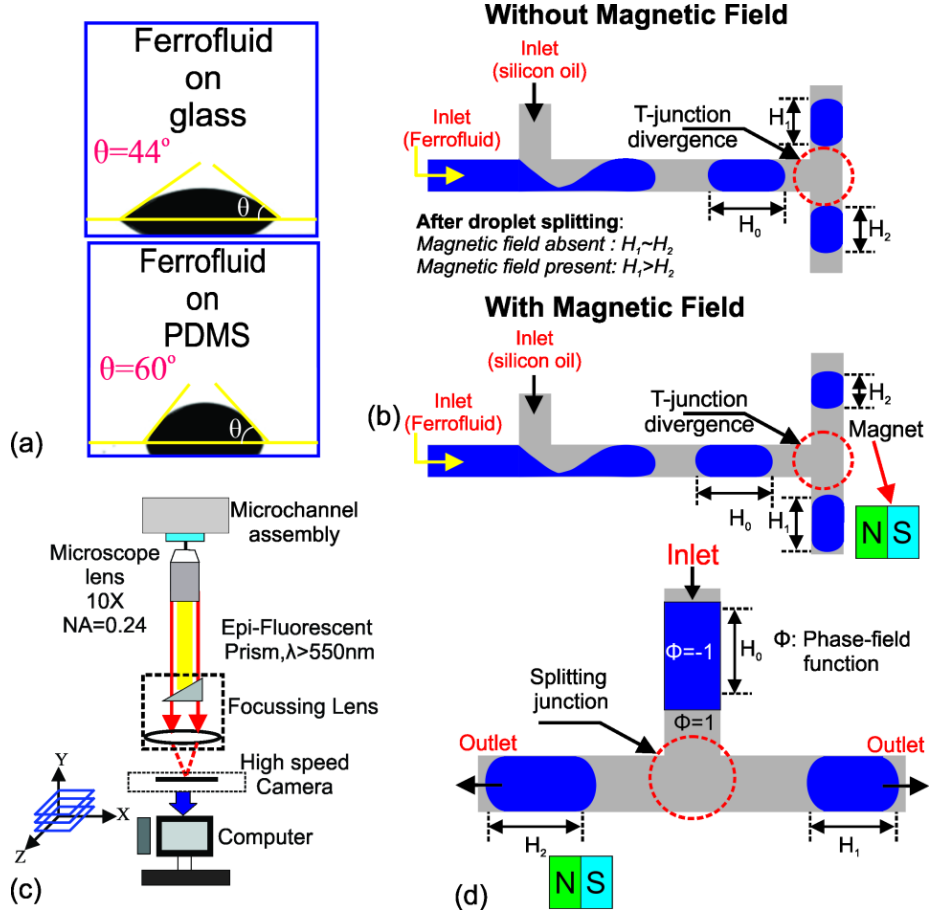

**FIGURE** 2. **(Color online)** (a) Contact angle of a typical ferrofluid droplet in a glass and PDMS substrate. Ferrofluid droplet has a contact angle($\theta$) of $44°$ and $60°$ on a glass and PDMS substrate, respectively, (b) Schematic representation of the working mechanism of the proposed microfluidic platform for controlled droplet splitting. Droplets on getting generated in a T-junction break down in a T-junction divergence under the influence of a non-uniform magnetic field, leading to its asymmetric splitting. The sequence of operation in the presence and absence of a magnetic field is schematically represented, (c) Graphical representation of the inverted microscope in which the experiments are carried out, (d) Schematic of the numerically simulated two-dimensional computational domain.

We use a Gaussmeter (Make: SES instruments) for the measurement of the magnetic flux density. The experiments are conducted in an inverted microscope (Make: Leica), schematic of which is shown in Figure 2(c). We employ high-speed imaging for recording the droplet splitting phenomena. We capture images of resolution $1920 \times 1280$ pixels$^2$ at a frame rate of 1000 fps. The recorded images are further post-processed in Matlab® platform by using an in-house developed code. Interested readers are referred to one of our previous works wherein the involved steps of the images processing have been discussed elaborately (Shyam *et al.* 2020b).

## 2.2. NUMERICAL ANALYSIS

### 2.2.1 System description

We also take an effort to investigate the droplet splitting phenomena numerically, essentially to understand the several intricate physical aspects involved with the underlying phenomena. Although, the experimental flow dynamics pertains to that of a droplet train flow scenario, through the numerical simulations, the dynamics of an isolated droplet in an immiscible liquid medium are explored. This particular exercise saves computational time significantly. Moreover, the prime intention of carrying out numerical simulation is to explore the isolated droplet breakup dynamics in the presence of a non-uniform magnetic field; as such, the inferences can be extrapolated to the present experimental scenario without compromising with the flow physics of our interest. In Figure 2(d), we show the schematic of the computational domain consisting of a symmetric T junction divergence and an isolated droplet surrounded by the continuous liquid. The isolated dispersed phase is driven forward by the continuous liquid. The droplet experiences breakage at the T-junction divergence. At the downstream of the T-junction (left branch), the non-uniform magnetic field is applied, as can be observed from Figure 2(d). The physical properties of the dispersed (denoted by suffix $d$) and continuous (denoted by suffix $c$) fluids are density, $\rho_d, \rho_c$; the viscosity, $\eta_d, \eta_c$; the magnetic permeability, $\mu_d, \mu_c$.

### 2.2.2 Phase-field formalism

We employ the diffused interface-based phase-field method for simulating the flow dynamics of immiscible two-phase flow systems as considered in this study. It may be mentioned here that the modeling framework of phase-field method is obtained by the minimization of the total energy of the system and is thermodynamically consistent (Cahn & Hilliard 1958). The phase-field method has been successfully used by several researchers to study the interfacial dynamics of a multi-fluid system in the presence of an external force field (Gorthi *et al.* 2017; Mondal *et al.* 2013, 2015). The thermodynamics of a two-fluid flow system can be described by the Ginzburg–Landau free energy functional $F(\phi)$, which is expressed as (Jacqmin 2000),

$$F(\phi) = \int_\forall \{f(\phi) + \frac{1}{2}\gamma\epsilon|\nabla\phi|^2\}d\forall \tag{1}$$

where $\forall$, spans over the whole fluid domain, $\gamma$, and $\epsilon$ are the interfacial tension and the interface thickness, respectively. The first term in Eq. (1) denotes the bulk free energy of the binary fluid system, while the second term signifies the interfacial free energy due to the presence of an interface separating the two fluids. The bulk free energy density $(f(\phi))$ can be expressed as, $f(\phi) = \gamma(\phi^2 - 1)^2/4\epsilon$. It may be mentioned here that the maxima and minima of $f(\phi)$ corresponds to the two stable phases involved, i.e., the dispersed phase $(\phi = -1)$ and the continuous phase $(\phi = 1)$. The minimization of the free energy of the system $(F(\phi))$ along with the mass conservation of the respective phases, leads to the well-known Cahn-Hilliard equation. Moreover, the minimization of the free energy of the system also leads to the addition of a volumetric force term in the Navier-Stokes equation, as will be discussed in section 2.2.4. The spatio-temporal evolution of $\phi$ depends on the Cahn-Hilliard equation as given by (Badalassi *et al.* 2003; Cahn & Hilliard 1958, 1959),

$$\frac{\partial \phi}{\partial t} + \boldsymbol{u}.\nabla \phi = \nabla.\left(M_\phi \nabla G\right) \tag{2}$$

where $M_\phi$ and $G$ denotes the interfacial mobility factor and chemical potential respectively. $M_\phi$ determines the relaxation time of the interface and the time scale of the Cahn-Hilliard diffusion. While the chemical potential $(G)$ is basically the variational derivative of the free energy functional with respect to the order parameter $(\phi)$ (Badalassi *et al.* 2003),

$$\begin{aligned} G &= \delta F/\delta \phi = \partial f/\partial \phi - \gamma \epsilon \nabla^2 \phi \\ &= \gamma[(\phi^3 - \phi) - \epsilon^2 \nabla^2 \phi]/\epsilon \end{aligned} \tag{3}$$

It may also be mentioned that in the phase-field framework, any generic property, $\xi$ may be expressed in terms of the order parameter $(\phi)$, as follows (Badalassi *et al.* 2003)

$$\xi = \frac{1-\phi}{2}\xi_c + \frac{1+\phi}{2}\xi_d \tag{4}$$

### 2.2.3 Modeling of Magnetic Field Distribution

We calculate the magnetic field acting on the flow domain by solving the Maxwell equations as given by (Griffiths 2017):

$$\nabla \cdot \boldsymbol{B} = 0 \tag{5}$$

$$\nabla \times \boldsymbol{H} = 0 \tag{6}$$

where $\bar{B}$ is the magnetic flux density and $\bar{H}$ is the intensity of the magnetic field. The magnetic flux density $(\bar{B})$ is given by (Griffiths 2017):

$$\bm{B} = \mu_0(\bm{H} + \bm{M}) \tag{7}$$

$\mu_0 = 4\pi \times 10^{-7}\ H/m$ is the permeability of vacuum, $\bm{M}$ is the magnetization vector. Since the magnetic field is irrotational (i.e., $\nabla \times \bm{H} = 0$), it can be expressed in terms of scaler potential, $\bm{H} = -\nabla \psi$. Therefore, using the scaler potential, Eq. (7) can be written as $\nabla \cdot (\mu \nabla \psi) = 0$. In the presence of a magnetic field, the total magnetic force that acts on the fluid volume is given by (Strek 2008):

$$\bm{F_m} = (\bm{M} \cdot \nabla)\bm{B} \tag{8}$$

### 2.2.4 Coupling of phase-field and magnetohydrodynamics

We solve the continuity equation, the Cahn Hilliard, and Navier-Stokes equations to obtain the pressure and velocity field of the two-liquid flow system, which are expressed as (DasGupta *et al.* 2014; Jacqmin 1999, 2000; Mondal *et al.* 2015),

$$\nabla \cdot \bm{u} = 0 \tag{9}$$

$$\rho \left[\frac{\partial \bm{u}}{\partial t} + \nabla \cdot (\bm{u} \cdot \bm{u})\right] = -\nabla \mathrm{P} + \nabla \cdot [\eta \nabla \bm{u} + (\nabla \bm{u})^T] + G\nabla \phi + \bm{F_m} \tag{10}$$

The momentum transport equation (Eq. (10)) couples phase-field formalism with magnetohydrodynamics. Here, $G\nabla\phi$, is the phase-field dependent interfacial tension force, $\bm{F_m}$ is the magnetic body force as denoted by Eq. (8). The boundary conditions considered in the flow domain for the simulations are as follows: fully developed velocity is applied at the inlet. Non-viscous, pressure constraint outflow condition is maintained at the outlet; in other words, the pressure at the outlet is fixed to zero. i.e., $P = 0$. For the magnetic field simulations, the magnetic insulation boundary condition, i.e., $\bm{n} \cdot \bm{B} = 0$, is applied at the surrounding air domain.

### 2.2.5 Normalization of the governing equations

In the present section, the aforementioned governing equations are non-dimensionalized for attaining a profound understanding. The width $(l)$ of the microchannel is considered as the characteristics length scale of the flow. The following set of dimensionless variables are defined for normalization of the governing equations:

$$u^* = \frac{ul^2}{Q},\ t^* = \frac{tQ}{L^3},\ P^* = \frac{Pt^4}{\rho_c Q^2},\ \rho^* = \frac{\rho}{\rho_c},\ \eta^* = \frac{\eta}{\eta_c},\ \mu^* = \frac{\mu}{\mu_c},\ \nabla^* = l\nabla,\ H^* = \frac{H}{H_0}$$

where $Q, \mu$, and $P$ denotes the flow rate, relative permeability, and pressure, respectively. Using the above set of non-dimensionalized variables, the governing equations i.e. the Cahn Hilliard, continuity, and Navier-Stokes equation gets reduced to

$$\frac{\partial \phi}{\partial t^*} + \boldsymbol{u}^* \cdot \boldsymbol{\nabla}^* \phi = \frac{3}{2\sqrt{2}} \frac{1}{Pe} \boldsymbol{\nabla}^{*2} [\phi(\phi^2 - 1) - Cn^2 \boldsymbol{\nabla}^{*2} \phi] \tag{11}$$

$$\boldsymbol{\nabla}^* \cdot \boldsymbol{u}^* = 0 \tag{12}$$

$$\rho^* \left[ \frac{\partial \boldsymbol{u}^*}{\partial t^*} + \boldsymbol{\nabla}^* \cdot (\boldsymbol{u}^* \cdot \boldsymbol{u}^*) \right] \tag{13}$$

$$= -\boldsymbol{\nabla}^* P^* + \frac{1}{Re} \boldsymbol{\nabla}^* \cdot [\eta \boldsymbol{\nabla}^* \boldsymbol{u} + (\boldsymbol{\nabla}^* \boldsymbol{u})^T]$$

$$+ \frac{3}{2\sqrt{2}} \frac{1}{CnCaRe} [\phi(\phi^2 - 1) - Cn^2 \boldsymbol{\nabla}^{*2} \phi] \boldsymbol{\nabla}^* \phi$$

$$+ \frac{Bo_m}{Ca} (\boldsymbol{M} \cdot \boldsymbol{\nabla}^*) \boldsymbol{B}$$

Thus the present problem of two-liquid flow systems is characterized by the following set of dimensionless parameters:

$$Re = \frac{\rho_c Q}{L \eta_c}, \ Cn = \frac{\epsilon}{L}, \ Ca = \frac{\eta Q}{L^2 \gamma}, \ Pe = \frac{\epsilon Q}{ML\gamma}, \ Bo_m = \frac{\mu_0 \chi l H^2}{\gamma}$$

The Reynolds number, $Re$, is the ratio of the inertia force to the viscous force. The Cahn number, $Cn$, is the ratio of interface thickness to the characteristics length. The capillary number, $Ca$, is the ratio of the viscous force to the interfacial force. The phase-field Peclet number, $Pe$, is the ratio of the advection of the order parameter ($\phi$) to its diffusion. The magnetic Bond number, $Bo_m$, is the ratio of the magnetic force to surface tension force ($\chi$ is the magnetic susceptibility). It is worth mentioning here that for the present analysis, the order of several non-dimensional numbers are considered as, $Re \sim O(10^{-3})$, $Ca \sim O(10^{-2})$, $Cn \sim O(10^{-2})$, $Pe \sim O(10^2)$ (Kunti *et al.* 2018; Mondal & Chaudhry 2018). The non-dimensional form of the fluid properties are given as (Badalassi *et al.* 2003)

$$\rho^* = \frac{1 - \phi}{2} + \frac{1 + \phi}{2} \rho_r \tag{14}$$

$$\eta^* = \frac{1 - \phi}{2} + \frac{1 + \phi}{2} \eta_r \tag{15}$$

$$\mu^* = \frac{1 - \phi}{2} + \frac{1 + \phi}{2} \mu_r \tag{16}$$

Note that $\rho^*, \eta^*, \mu^*$ in Eq. (14)-(16) denotes the ratio of the respective properties of the dispersed phase to that of the continuous phase liquid. Similarly, $\rho_r, \eta_r, \mu_r$ in Eq. (14)-(16) indicates the ratio of the properties of the continuous phase with that of the dispersed phase. For the present numerical study: $\rho^* \sim O(10^0)$, $\eta^* \sim O(10^{-2})$ and $\mu^* \sim O(10^1)$. The unsteady governing transport equations together with the boundary conditions in the present study are solved by using the finite element framework of COMSOL Multiphysics®. In this method, the spatial terms of the governing transport equations are initially discretized to obtain the ordinary differential equation in time, which is then time marched to obtain evolution of the flow pattern. The PAR solver with generalized-$\alpha$ scheme is used for time-stepping method. The Galerkin Weighted Residual Method, blended with higher-order up-winding and stabilized with the capability of handling crosswind diffusion, has been deployed here for the discretization of the convection-diffusion equations. The second-order interpolation function for velocity and the first-order interpolation function for pressure are used for the calculations of gradients within the integral terms under the framework of the weak formulation. The interpolation function for pressure is one order lower than that used for velocity. In order to obtain the velocity and pressure profiles for the incompressible flows, the incremental pressure correction scheme for the segregated predictor-corrector method is employed. The tolerance levels of $10^{-6}$ for mass divergence are specified for all the numerical simulations (Mondal *et al.* 2015).

**2.2.6 Grid independence study and Validation:**

Here, we show the grid independence test to ensure the correctness of the numerical results presented in this analysis. In addition to that, to ensure a sharp interface limit, we also carry out the Cahn number independence test. It may be mentioned here that in our study, we have chosen the grid size near the interface equal to the Cahn number ($Cn$). Thus, in the context of the present analysis, a grid-independence will simultaneously indicate a Cahn number independence and vice-versa. Figure 3(inset) shows the temporal evolution of the non-dimensionalized width ($W^*$) of the droplet as it breaks up in the T junction divergence of the LOC device, obtained for different values of grid resolution. As can be seen clearly from Figure 3(a) that the numerical results become independent of mesh size below $Cn < 0.02$. Although for $Cn \sim 0.01$, a better resolution is obtained, however, considering the involved computational cost vis-a-vis the gained accuracy, we choose $Cn \sim 0.02$ for the present study. Note that this particular value of Cahn No, $Cn \sim O(10^{-2})$ ensures the sharp interface limits as well (Yue *et al.* 2010).

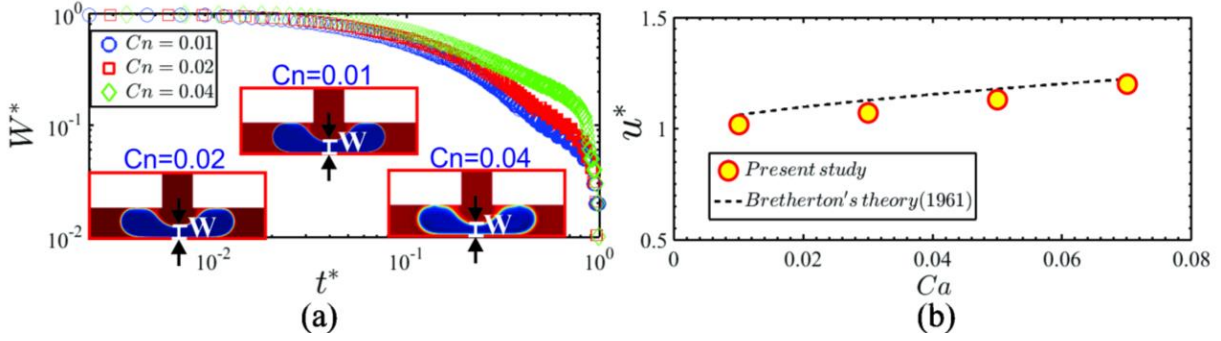

**FIGURE** 3. **(Color online)** (a) Temporal evolution of the non-dimensionalized width ($W^*$) of the droplet at the T-junction divergence for different grid resolutions. Grid independence test justifies Cahn number independence test. $W^* = w(t)/l$, where $w(t)$ and $l$ denotes the instantaneous width of the droplet and the width of the microchannel, respectively. Inset snapshots shows a typically resolved interface of a droplet for the Cahn number, $Cn = 0.01, Cn = 0.02,$ and $Cn = 0.04$, (b) Plot shows the comparison between the present numerical study and Eq. (17).

We also compare in Figure 3(b) our numerical results quantitatively with those obtained from the analytical relation of Bretherton (Bretherton 1961). For low Reynolds number flow i.e., $Re \ll 1$, the velocity of a droplet moving in a slender tube, with a thin film separating the droplet and the wall is given by (Bretherton 1961),

$$u^* = 1 + 1.29 Ca^{2/3} \qquad (17)$$

We show in the inset of Figure 3, an excellent agreement between our numerical results with those obtained from Eq. (17). A closer match between the results, seen from Figure 3(b), vouches for the correctness of the numerical modelling framework developed in this study.

## 3. RESULTS AND DISCUSSIONS

In this section, we explore the droplet breakup phenomena and its consequences to the alterations in the various hydrodynamical parameter of the flow field. As already mentioned, the present investigation is carried out in the flow regimes whereby the viscous and interfacial forces are dominant in comparison to the inertia forces. As such, the study is divided into four parts. In the first part, we develop a physical understanding of the ferrofluid droplet splitting mechanism in the presence of a non-uniform force field. To this end, we make use of the high-speed imaging modalities and perform the numerical simulations for exploring the dynamical behavior of the droplet under the modulation of the spatially varying force field. In the second section, we numerically explore the intricate flow physics involved with the droplet splitting dynamics. Subsequently, in the third section, we experimentally explore the morphological evolution of the mother droplet in the T-junction divergence of the LOC-device under the modulation of the non-uniform magnetic forcing. Finally, in the last part of the study, we make an attempt to develop a physical reasoning behind the characteristic behavior of the sister droplets from the perspective of the involved time scales.

### 3.1 Droplet Break up: Qualitative Dynamics

It may be mentioned here that depending on the capillary number $(Ca)$, and initial slug length $(l_0)$, a droplet may split into two sister droplets in the T-junction divergence following permanent obstruction or partial obstruction (Jullien *et al.* 2009; Leshansky & Pismen 2009). In addition to that, based on the initial slug length and Capillary number, the droplet may not split at all (Chen & Deng 2017; Jullien *et al.* 2009). Pertaining to the permanent obstruction case, the dispersed phase totally engulfs the T-junction divergence and blocks the flow of continuous phase liquid. On the contrary, for the partial obstruction case, a tunnel develops, which allows the motion of the continuous fluid over the dispersed phase. Therefore, in the partial obstruction case, the droplet blocks the T junction divergence of the LOC-device incompletely. Henceforth, the word 'droplet' will refer to the mother droplet.

It is worth mentioning here that for the present study, the mother droplet splits into two sister droplets following the permanent obstruction of the T-junction divergence, as can be observed from the depicted images in Figure 4. We show in Figure 4 the spatio-temporal evolution of the droplet splitting phenomena in the absence of an external field. Note that droplet breaks ensuing three typical stages: squeezing, transition, and pinch-off (Ma *et al.* 2017). Initially, as the droplet (i.e., dispersed phase) enters into the T junction divergence, it tries to block the whole junction, as can be seen from Figure 4(a). As the droplet occupies the whole junction $(t^* = 0^+)$, the upstream continuous phase flow gives rise to the formation of a depression in the droplet, which further, leads to the change in the curvature of the neck (the circled region in $t^* = 0.5$ in Figure 4(a)). The depression is a resultant effect of the squeezing pressure acting on the neck of the droplet (of the dispersed phase). The squeezing pressure is the pressure that is being developed in the upstream continuous phase due to the permanent blockage of the T-Junction divergence by the dispersed phase. Since no tunnel develops for the present case (i.e., permanent obstruction case), the squeezing force becomes very large in comparison to the viscous force, thereby dictating the overall dynamics in the squeezing stage. The squeezing stage is followed by the transition stage. In the transition stage, too, no tunnel formation is seen, and thus, the splitting phenomena is dictated by the balance between the interfacial tension force and the squeezing force. Although the upstream pressure force (i.e., the squeezing force) is dominant in both the squeezing and transition stage, the temporal evolution of the rear interface of the droplet is found to be different for the two individual cases, as will be discussed in detail in the forthcoming sections. The transition zone is followed by the pinch-off zone, in which the droplet gets fully detached, leading to the formation of two sister droplets, as can be observed from Figure 4(a). In the succeeding discussions, we will demarcate all of these existing splitting zones appropriately.

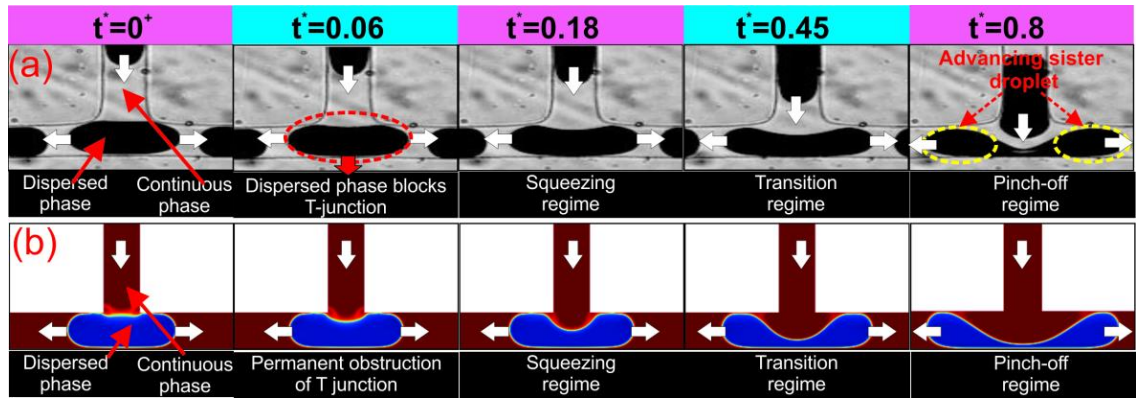

**FIGURE** 4. **(Color online)** Temporal evolution of the droplet splitting phenomena observed using (a) Experiments, and (b) Numerical simulations. The white-colored arrows indicate the direction of the flow. $t^* = t/t_L$ represents the non-dimensional time period of the droplet splitting phenomena. $t, t_L$ represents the instantaneous time and the droplet splitting time, respectively. The span of the droplet splitting time starts from the moment the droplet fully enters the T-junction to the moments two dispersed sister droplet has been formed. The experiments are carried out using at a magnification of 10X at a capturing speed of 1000 frames per second.

To attain a detailed visualization alongside to arrive at an in depth understanding of the experimental observation, we perform numerical simulations of the droplet splitting phenomena. The details of the numerical methodology adopted in this analysis is already mentioned in section 2.2. As already mentioned before that although the present experimental study deals with droplet train, the splitting dynamics of an isolated droplet is investigated numerically. This particular exercise will give us qualitative insights into the droplet breakup dynamics in the absence/presence of a non-uniform magnetic field. We show in Figure 4(b) the droplet splitting phenomena in a T-junction simulated numerically. We can clearly observe the presence of various droplet splitting stages such as the squeezing, transition, and pinch-off from Figure 4(b). It is worth to add here that the similarity of the underlying droplet splitting dynamics between experimental observation and numerical results, observed from qualitative perspectives in Figure 4(a) vis-à-vis Figure 4(b) justifies the credibility of our experimental methods.

We show in Figure 5(a)-(b) the spatio-temporal evolution of the ferrofluid droplet splitting phenomena in the presence of a non-uniform magnetic field, obtained experimentally and calculated from numerical simulations, respectively. The presence of all the intermediate stages of the splitting phenomena, i.e., *the squeezing stage, transition stage, and the necking stage,* which are observed in Figure 4 as well, can be clearly seen for the present case (see Figure 5). The behavior of the droplet splitting in the presence of a non-uniform magnetic field becomes similar to the case where no external force acts on the droplet flow field. The only difference is the asymmetricity produced during the droplet breakup process. Note that the asymmetricity developed in the droplet splitting process is precisely due to the involved non-uniformity in the force field gradient induced due to the applied field. Readers are referred to the supplementary materials section for a detailed distribution of the magnetic field flux density in the fluid flow domain. We can clearly observe from Figure 5(a), that as the ferrofluid droplet enters the T-junction divergence, the upstream pressure forces the advancement of the dumbbell shaped bulge in the left and right branches, respectively. Note that in the succeeding discussion, we refer these left/right dumbbells shaped bulges of the dispersed phase as the left/right moving bulge.

However, due to the non-uniform distribution of the applied magnetic flux density, the advancement (of the dumbbell-shaped bulge) becomes more aligned towards the left branch (asymmetricity is produced in the break-up phenomena) in comparison to the right branch (refer $t^* = 0.5$ and $t^* = 0.75$ of Figure 5(a)) of the T-junction divergence. Particularly, because of this reason, we observe in Figure 5(a) the presence of unevenly sized sister droplets being produced in the process. As such, the larger-sized sister droplet moves in the left branch, while the smaller-sized sister droplet moves in the right branch, as can be clearly observed from Figure 5(a). We show the numerically simulated droplet splitting phenomena in Figure 5(b). The asymmetric behavior of the droplet breakup can easily be observed from Figure 5(b) as well. Quite notably, the numerically simulated stages of the break-up phenomena are in accordance with the experimental observations (see figure 5(a)), further justifying the credibility of our experimental observation. Moreover, it may be conjectured from Figures 4-5 that for a droplet splitting following permanent obstruction, the characteristic behavior of the breakup pattern remains the same irrespective of the fact whether a magnetic field is applied or not. The utilization of a non-uniform magnetic field gives rise to asymmetry in the droplet breakup phenomena, leading to the generation of two sister droplets of unequal sizes.

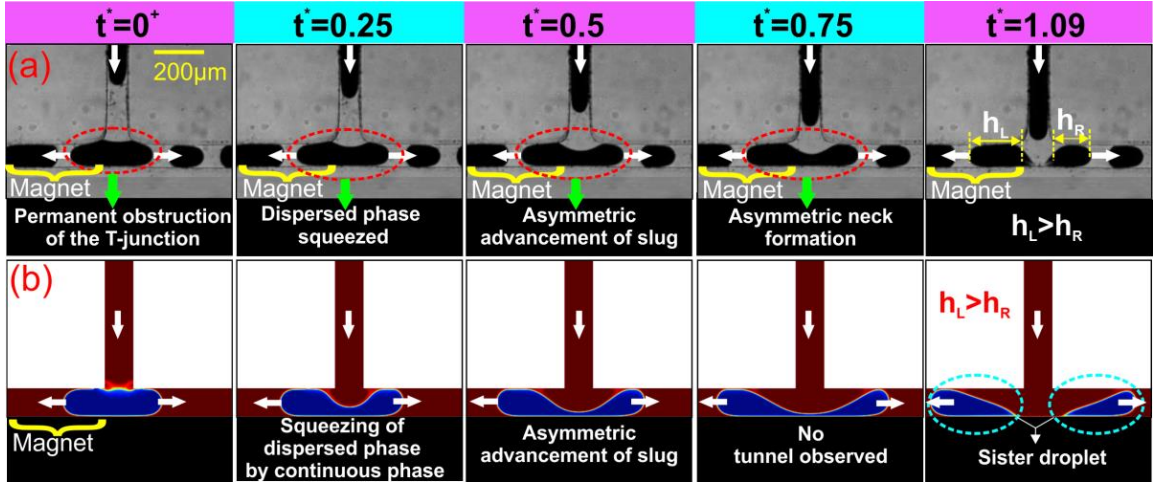

**FIGURE** 5. (**Color online**) Temporal evolution of the droplet splitting phenomena in the presence of a non-uniform magnetic field, observed using (a) Experiment, (b) Numerical simulations. The white-colored arrows indicate the direction of the flow. $t^* = t/t_L$ represents the non-dimensional time period of the droplet splitting phenomena. $t, t_L$ represents the instantaneous time and the droplet splitting time, respectively. The span of the droplet splitting time starts from the moment the droplet fully enters the T-junction to the moments two dispersed sister droplet has been formed. The experiments are carried out using a higher magnification of 10X at a capturing speed of 1000 frames per second.

### 3.2 Droplet Breakup Phenomena: Numerical Perspectives

As seen from the preceding discussion that the numerical results agreed well with the experimental observations. This endeavour provides us the flexibility to delve deep into the droplet splitting dynamics following numerical computations. We show in Figures 6(a)-(b), the velocity vectors inside the droplet flow field at different stages of splitting for the various cases under investigation, i.e., in the absence/presence of a non-uniform magnetic field. It may be mentioned here that $t^* = 0^-$ denotes the stage of the droplet before entering the T-junction divergence. It can be clearly observed that the center of the droplet before fully impacting the T-junction divergence attains maximum velocity (see $t^* = 0^-$ of Figure 6(a)). In the absence of any external force, we can clearly observe that the droplet demonstrates two distinct velocity regimes when it impacts the T-junction divergence at $t^* = 0^+$, as denoted by point **A** and point **B** in Figure 6(a).

Here, point **A** identifies the position whereby the droplet remains in contact with the upstream flow, point **B** on the otherhand, denotes the position of the droplet which faces the wall of the T-junction divergence. Note that point **A** demonstrates the maximum velocity while point **B** depicts the region of minimum velocity. This is particularly because the upstream flow attempts to squeeze the droplet, thereby ensuring that point **A** exhibits higher velocity. Similarly, as the droplet impacts the wall of the junction, a stagnation point exists, as indicated by point **B** in Figure 6(a). Subsequently, the droplet, after impacting the T-junction divergence, stretches itself symmetrically in the left and right branches, respectively. This stretching leads to the migration of the local maxima points, as given by points **C** (refer $t^* = 0.25$ of Figure 6(a)). On the other hand, the position of the local minima (point **B**) remains the same due to the symmetric nature of the stretching phenomena, as can be observed from Figure 6(a). It can also be observed that the magnitude of velocity inside the droplet decreases with the temporal progression of splitting phenomena, as witnessed in Figure 6(a). Having a closer look at Figure 6(a) vis-à-vis Figure 6(b), it is observed that the internal hydrodynamic behaviour inside the droplet under the influence of a non-uniform magnetic field gets totally altered. The position of the maximum and minimum velocity inside the droplet is represented by point **D** and **E.** It can be clearly seen from Figure 6(b) that the position of the local maxima and minima inside the droplet is altered when compared to the corresponding points pertaining to the non-magnetic case ($t^* = 0^+$ of Figure 6(b)). This altercation is due to the presence of the non-uniform magnetic flux density. The presence of high force field gradient in the left branch ensures that the maxima (of velocity) is always aligned towards the left branch. Consequently, due to low magnetic flux density, the minima (of velocity) is aligned towards the right branch as can be clearly observed from Figure 6(b). In particular, a high force field gradient acting in the left branch induces more ferrofluid mass to flow towards it, leading to the generation/development of large-sized sister droplets moving on the left side (refer to Figure 6(b)). Also, the location of the stagnation point is altered due to the involved asymmetric stretching of the droplet (refer point **E** of Figure 6(b)). Quite notably, the influence of a magnetic field leads to the migration of the localized maxima and minima to the left bulge (of the droplet) as can be seen at $t^* = 0.5$, and 0.75, respectively from Figure 6(b). In Figure 6(c), we show the zoomed-in view of the droplet domain at $t^* = 0.75$ for both the cases i.e., with/without magnet cases, essentially to draw clear distinctions between them. The presence of maxima and minima in the left moving bulge (of the droplet) under the modulation of a magnetic field can be clearly observed in Figure 6(c-II). Note that the fluid facing the lower wall (i.e., the wall of the left branch facing the magnet) has a low magnitude of velocity in comparison to the fluids facing the upper wall. As such, two clear distinct zones can be observed in the left moving bulge (of the droplet) under the modulation of a non-uniform magnetic field.

To explore the implication of these typical zones, in the Figure 6(c-III), we plot a further zoomed-in view of the left moving bulge of the dispersed phase (i.e. the droplet) under modulation of magnetic field (at $t^* = 0.75$). We demarcate the fluid velocity magnitude inside the left moving bulge into two zones i.e. the zone I and zone II, respectively, as shown by the pink colored ellipse in Figure 6(c-III). As clearly visible, the zone I, which is adjacent to the lower wall, experiences a velocity of lower magnitude. On the contrary, the zone II, which is adjacent to the upper wall, experiences velocities of higher magnitude comparatively. Note that the lower wall experiences higher magnetic field flux density in comparison to the upper wall. Therefore, it may be argued upon that when the ferrofluid droplet reaches the T-junction divergence, the magnetically susceptible fluid rushes towards the lower wall in an attempt to approach the magnet. However, on reaching the high magnetic field flux zone, the ferrofluids (inside the left bulge) motion is highly restrained. This restrained motion of the fluid towards the magnet leads to the development of a "*hump-like structures*" nearby the lower wall as can be observed from Figure 6(c-III). Due to the high magnetic field gradient existing nearby the lower wall, the magnitude of velocity is low inside this hump-like structure as can be clearly observed from Figure 6(c-III). It may also be mentioned here that owing to the presence of hump-like structure, the succeeding ferrofluids (inside the left moving droplet volume) are forced towards the upper wall by satisfying the mass conservation constraint, as can be observed in zone II from the right most snapshots of Figure 6(c-III). Due to the relative low force field gradient nearby the upper wall, high velocity magnitude is encountered nearby the upper wall as can be clearly seen from Figure 6(c-III). Following this, it may be argued that presence of the "*hump-like structure*" non-trivially ensures augmented convections inside the left-directed bulge of the droplet, thereby endorsing its amplified velocity in comparison to the bulge advancing in the right branch.

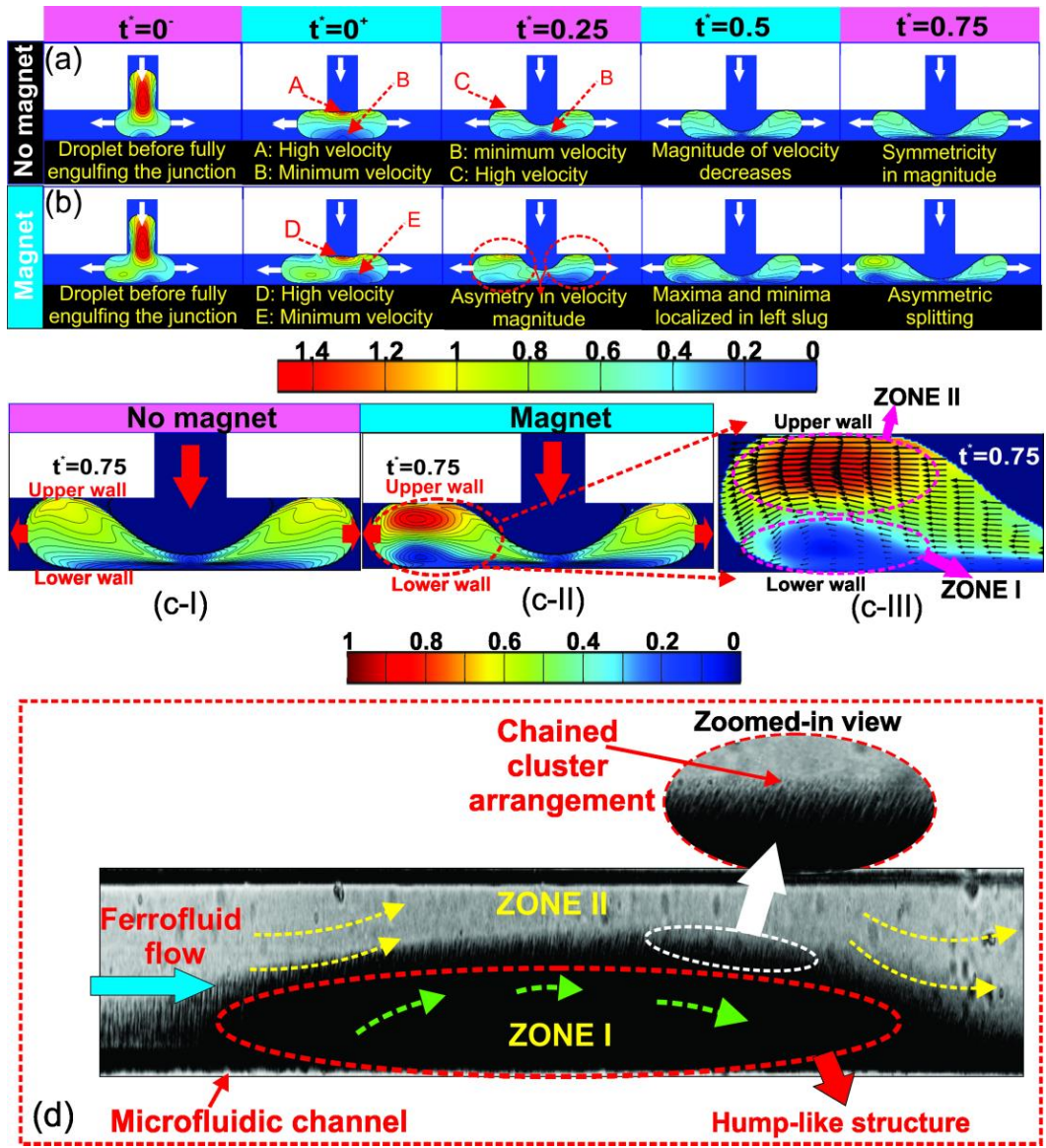

**FIGURE 6. (Color online)** Temporal evolution of the droplet splitting phenomena calculated numerically in (a) absence and (b) presence of a non-uniform magnetic field. The white-colored arrow indicate the direction of the flow. $t^* = t/t_L$ represents the non-dimensional time period of the droplet splitting phenomena. $t, t_L$ represents the instantaneous time and the droplet splitting time respectively, (c) Zoomed view of the droplet splitting phenomena at $t^* = 0.75$ for I. with and II. without magnet case. The red-colored arrow shows the direction of the flow. III. The right most figure shows the zoomed view of the red colored ellipse along with the black colored arrows indicating the velocity vectors, (d) Experimental hump formation in the ferrofluid flow domain in presence of a static magnet. The zoomed-in view shows the chained cluster formation of the magnetic nanoparticles in the hump-like structure. The yellow and red colored dotted arrow indicate the direction of the ferrofluids in zone I and II respectively.

In the previous discussion, we have argued upon the presence of "*hump-like structure*" inside the left moving ferrofluid droplet under the modulation of a magnetic field. To check the validity of this arguments, we carried out a bright field visualization of the ferrofluid flow field in a microchannel under the influence of a static magnet. We show in Figure 6(d), the ferrofluid flow field in the presence of a steady magnet placed nearby the lower wall. The presence of a hump-like cluster, together with the formation of nanoparticles agglomeration following chain-like structure, can be clearly observed from Figure 6(d). This observation justifies our arguments for the enhanced flow convections in the left moving bulge (see movie 4 for a clearer insights). We will show in the succeeding sub-sections that this augmented velocity in the bulge (moving in the left branch) can be significantly maneuvered in controlling the size of the generated droplet (precisely sister droplet) particularly, for the cases involving with the droplet train.

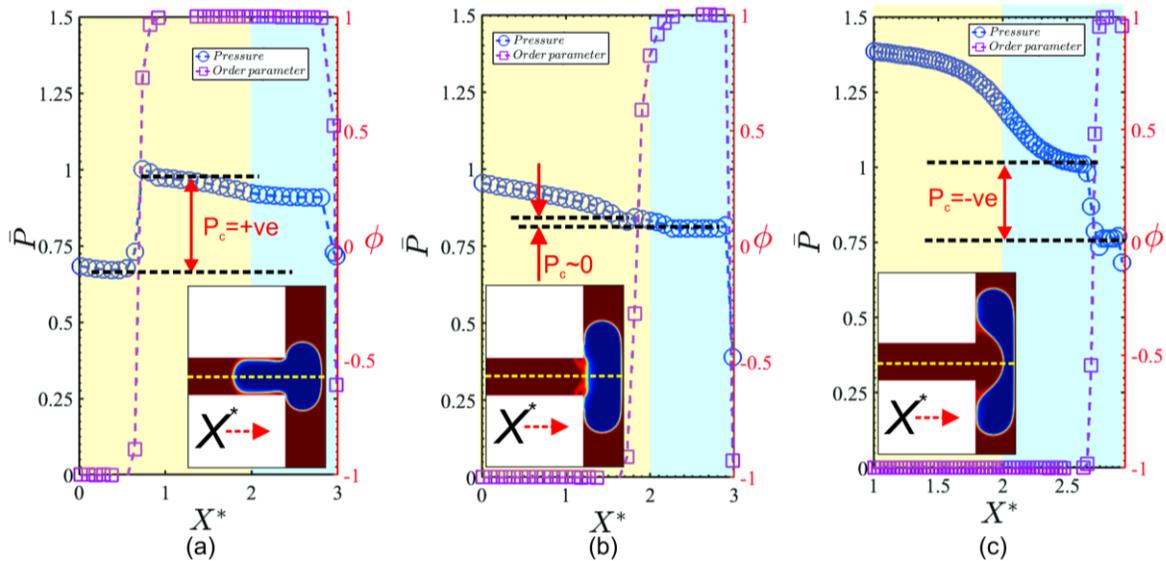

**FIGURE** 7. (**Color online**) Pressure and order parameter distribution along the symetric axis of the T-junction around the interface for the droplet in the (a) entering phase (b) at the junction (c) after entering the junction, in absence of any external force field; $P_c = P_{inside} - P_{outside}$, $P_c$ is the Laplace pressure, $P_{inside}$ is the pressure inside the droplet (dispersed phase), $P_{outside}$ is the pressure inside the upstream continuous phase, $\phi$ is the order parameter. $\bar{P} = P^*/P_{inlet}$, $P^*$ is the instantaneous non-dimensionalized pressure and $P_{inlet}$ is the non-dimensionalized pressure at the inlet of the channel. Negative value indicates surface tension is oriented upstream. The blue colour shaded area indicates the branched channel area. The inset indicates the respective spatio-temporal location of the droplet

As already discussed before in section (3.1) that for a droplet getting splitted into sister droplets following the permanent obstruction, the balance between the upstream pressure force, magnetic force, and the interfacial tension force dictates the overall splitting/break-up phenomena. Accordingly, in order to have a comprehensive understanding on the competition of these two forces, we show in Figures 7-8, the detailed evolution of the pressure distribution and the Laplace pressure drop across the rear droplet interface during the break-up phenomenon with permanent obstruction. The discussion pertaining to this aspect follows the results obtained from the numerical simulations performed in this analysis. It may be mentioned here that due to the upstream pressure, the curvature of the rear interface of the droplet undergoes a temporal change from convex to concave shape. Therefore, we represent the entire phenomena in Figures 7-8, by depicting three typical stages i.e., when the droplet rear interface is convex, flat, and concave, respectively (see inset of individual figure for the snapshot of the temporal instant). The order parameter distribution in Figures 7-8, helps in identifying the rear interface as it varies between $\phi = -1$ and $\phi = 1$. Accordingly, the pressure difference across the interface indicates the Laplace pressure $(P_c = P_{inside} - P_{outside})$.

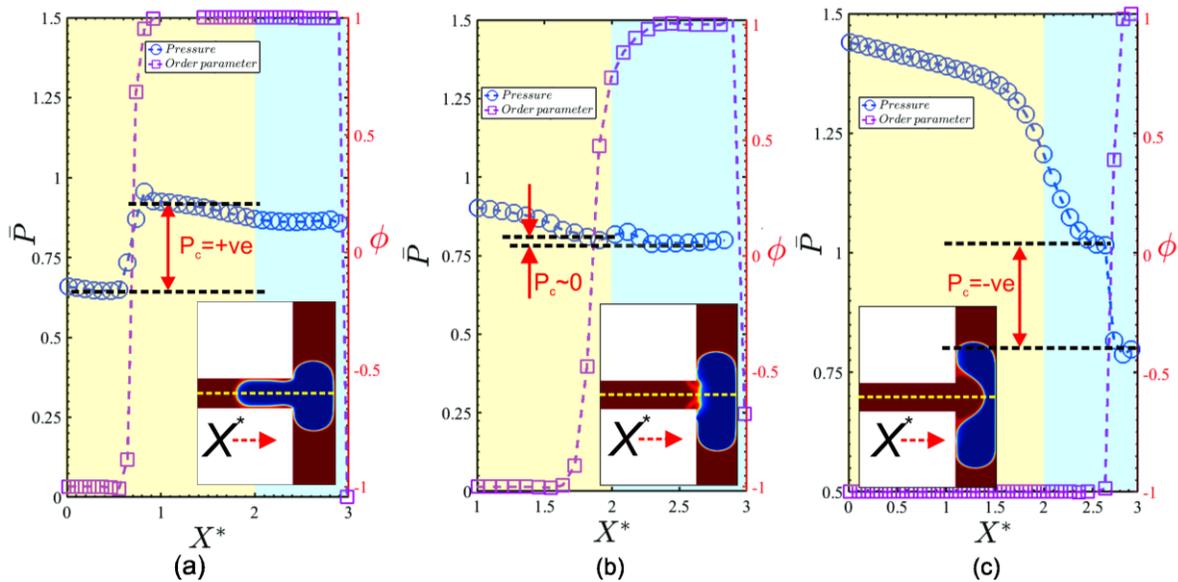

**FIGURE** 8. **(Color online)** Pressure and order parameter distribution along the symetric axis of the T-junction around the interface for the droplet in the (a) entering phase (b) at the junction (c) after entering the junction, in the presence of a non-uniform external magnetic field; $P_c = P_{inside} - P_{outside}$, $P_c$ is the Laplace pressure, $P_{inside}$ is the pressure inside the droplet (dispersed phase), $P_{outside}$ is the pressure inside the upstream continuous phase, $\phi$ is the order parameter. $\bar{P} = P^*/P_{inlet}$, $P^*$ is the instantaneous non-dimensionalized pressure and $P_{inlet}$ is the non-dimensionalized pressure at the inlet of the channel. Negative value indicates surface tension is oriented upstream. The blue colour shaded area indicates the branched channel area. The inset indicates the respective spatio-temporal location of the droplet

It may be mentioned here that in the absence of an external force field, when the droplet enters the T-junction, its motion becomes restrictive due to the presence of the wall of the junction. As a consequence of this restricted motion, the droplet undergoes deformation symmetrically. At this juncture, the rear interface of the droplet exhibits a convex profile, as can be observed from inset of Figure 7(a). Due to this convex profile, the capillary pressure, $P_c(=$ $P_{inside} - P_{outside}) \gg 0$, demonstrates a positive value (see Figure 7(a)). With the progression of temporal instances, the droplet leaves the main channel fully and occupies the whole junction, and at this stage, the upstream pressure forces the rear interface of the droplet to attain a flat profile ($t^* = 0$, of Figure 6(a)). It is due to this flat profile, the capillary pressure assumes almost a negligible value i.e., $P_c (= P_{inside} - P_{outside}) \sim 0$, as can be observed from Figure 7(b). The upstream pressure further forces the neck of the droplet to attain a concave profile, thereby ensuring that the capillary pressure attains a negative value i.e., $P_c (= P_{inside} - P_{outside}) < 0$, as can be observed from Figure 7(c). Note that the dynamical evolution of the pressure remains the same qualitatively even in the presence of a non-uniform magnetic field, i.e., the rear interface of the droplet evolves from the convex to the concave shape as observed from Figure 8(a)-(c). Up till this section, we correlate the insight gained from the numerical simulations directly to the experimental scenarios. Following these inferences, which will be used to support the laid down the experimental observations, we explore the dynamics of the droplet break-up events focusing aptly on experimental results in the succeeding sections.

### 3.3 Droplet Break up: Evolution of the Droplet Morphology

#### 3.3.1 Evolution of droplet width

We have seen from the previous discussion that on application of a non-uniform magnetic field, the ferrofluid droplet tends to get stretched in the direction of the applied magnetic field. We have also observed from both experimental observations and numerical simulation that the uneven stretching of interface results in an asymmetric splitting of the droplet. In Figure 9, we show the variation of the non-dimensional thickness ($W^* = w/l$) of the droplet (dispersed phase) as it breaks both in the presence and absence of a magnetic field. It may be mentioned here that the underlying event of droplet splitting is mainly governed by the intricate competition among the interfacial force, viscous force and the magnetic force. Here, we introduce the magnetic Bond number ($Bo_m$), which is used to represent the relative strength between the magnetic force and the surface tension force, respectively. The time zero indicates the moment at which the dispersed phase entirely penetrates into the T-Junction divergence. As such that the rear interface of the dispersed phase is almost flat at $t = 0$. We demarcate, in Figure 9, the various regimes of splitting encountered by the droplet i.e., *the squeezing regime, the transition regime and the pinch-off regime*. The demarcation of the various regimes is identified out following the characteristic slopes exhibited by the deforming ferrofluid droplet in the T junction divergence as can be observed from Figure 9. It may be mentioned here that evolution of the neck thickness of the droplet is such that in the initial *squeezing regime*, a linear gradient is observed. Important to mention here that this linear variation of $W^*$ is attributed to the typical role played by the two responsible forces i.e., the interfacial force and the upstream pressure force on the underlying phenomena in this regime. Due to the limited deformation of the rear interface of the droplet in the *squeezing regime*, the role played by the interfacial force in the pertinent regime becomes miniscule. As a result, the rear interface of the droplet moves primarily due to the squeezing force of the upstream flow, with limited to no role played by the surface tension force. It is because of this force balance, a linear relationship between $W^*$ vs $t$, can be observed during the squeezing regime as observed from Figure 9. This is followed by the *transition regime*, in which the interfacial tension force resists the deformation of the rear interface. This resistance leads to a delay in the overall deformation of the rear interface of the droplet. Consequently, we observe an exponential relationship between $W^*$ and $t$, as witnessed in Figure 9 (see regime B). The *transition regime* is followed by the *pinch off stage,* in which the thinning rate gets significantly amplified with time.

The top inset of Figure 9 shows the variation of the thinning rate, $1 - W^*$ vs $t^*$, for the case when magnetic field is applied. As already mentioned, in the squeezing regime, it can be observed, $1 - W^* \sim t^*$, and it is essentially due to the minimal involvement of interfacial tension force. While for the transition regime, we observe, $1 - W^* \sim t^{*3/7}$, where scaling 3/7, agrees well with the theoretical solutions as proposed by Leshansky and Pismen (Leshansky & Pismen 2009). Therefore, it can be argued that the characteristics behavior of the droplet splitting even, as observed in the regime of permanent obstruction, is independent of the fact weather a magnetic field is applied or not. However, it can be clearly visualized from Figure 9 that magnetic field has a role to play on the overall life time of the droplet break up phenomena. This further implies that by specifically tuning the force field, we could effectively control the size of the droplet. In the forthcoming sections, we comprehensively discuss effective ways in which the size of the sister droplets can be controlled.

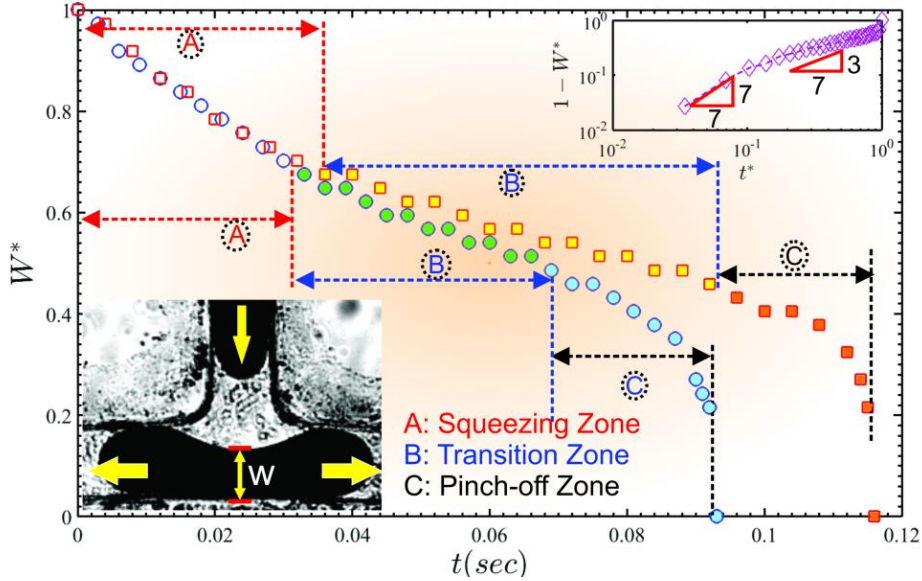

**FIGURE** 9. **(Color online)** Temporal evolution of the non-dimensionalized droplet width ($W^*$) in absence and presence of magnetic field. The circles indicate the case when $Bo_m = 0$, the squares indicate the case when $Bo_m = 6.1$. The blue, black, and red lines demarcates the squeezing, transition and pinch-off regimes. The inset shows the schematic of the width of a typical droplet breakup process under permanent obstruction case. $W^* = w(t)/l$, where $w(t)$ and $l$ denotes the instantaneous width of the droplet and the width of the microchannel. $R = 1$, $Ca = \sim 10^{-3}$, $Re \sim 10^{-3}$.

### 3.3.2 Effect of Magnetic Flux Density

We show in Figure 10, the variation of non-dimensionalized width ($W^*$) for the various magnetic flux densities (precisely, for various $Bo_m$). It can be clearly observed from Figure 10 that the characteristic variation of the droplet width is same irrespective of the applied magnetic field strength. Also, Figure 10 demonstrates that the required time of splitting of the ferrofluid droplet into the sister droplets is dependent on $Bo_m$. As such, it could be observed from Figure 10 that the droplet splitting time ($t$) varies in the following sequence: $Bo_m(= 0) < Bo_m(= 7.5) < Bo_m(= 6.1) < Bo_m(= 15.1)$. In the next sections, we explore this particular aspect of the droplet splitting time as modulated by the magnetic flux density in a greater detail.

The top right inset of Figure 10 shows the variation of non-dimensionalized length ($l_l^*$) of the sister droplet flowing in the left branch. Important to mention here that the magnet is placed adjacent to the left branch of the T-junction divergence. Therefore, to explore the role of magnetic field on the droplet splitting and subsequently, on the change in the length of the sister droplets, we focus our attention on $l_l^*$. Note that $l_l^* > 0.5$, ensures that the length of the sister droplet moving to the left branch ($l_l^*$) is more in comparison to the length of the sister droplet moving to the right branch ($l_r^*$). In other words, it can also be said that the application of non-uniform magnetic field ensures that an increased fluid volume moves to the left branch when compared to the volumes of fluid moving to the right branch, precisely due to the existing the high field gradient (in the left branch). We can clearly observe from Figure 10(right-inset) that the curve exhibits a positive slope and on reaching a particular threshold value, it exhibits a negative slope. Note that with an initial increase in $Bo_m$, the change in length of the sister droplet in left branch ($l_l^*$) increases and on reaching a particular critical $Bo_m$, the length of the sister droplet ($l_l^*$) decreases. Any increase in $Bo_m$, beyond the critical value reduces the size of the sister droplet moving in the left branch. The left-side inset (of Figure 10) shows the corresponding snapshots of the droplet splitting phenomena for the different cases under investigation. It can be clearly observed from the left-side inset of Figure 10 that the length of the sister droplet ($l_l^*$) is maximum for $Bo_m = 7.5$.

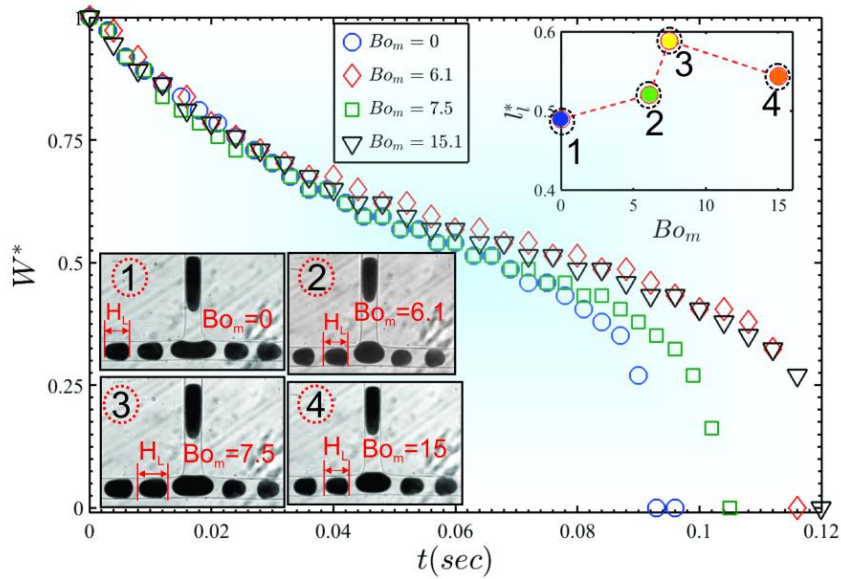

**FIGURE** 10. **(Color online)** Temporal evolution of the non-dimensionalized droplet width ($W^*$) for the various magnetic field Bond number ($Bo_m$). $W^* = w(t)/l$, where $w(t)$ and $l$ denotes the instantaneous width of the droplet and the width of the microchannel. The right side inset shows the variation of the non-dimensionalized length ($l_l^*$) of the sister droplet flowing in the left branch of the microchannel, for the various magnetic field strength. The corresponding snapshots of the generated droplet is shown in the left-bottom inset marked by 1, 2, 3, 4. $l_l^* = l(t)/l_0$, where $l(t)$ and $l_0$ denotes the instantaneous length of the sister droplet and characteristic length of the mother droplet. $R = 1$, $Ca = \sim 10^{-3}$, $Re \sim 10^{-3}$.

### 3.3.3 Effect of Flow Ratio ($R$)

In this subsection, we explore the effect of the dispersed phase flow rate on the droplet splitting phenomena. We have mentioned that in the present study the flow rate of the dispersed phase is changed while that of the continuous flow is kept constant. This particular exercise is carried out essentially to vary the length of the generated droplet (equivalently the volume), keeping the continuous phase flow velocity constant. We normalize the flow rate as flow ratio, given as, $R = Q_c/Q_d$. Note that lower the value of $R$, higher is the slug length of the dispersed phase (i.e., the droplet). We show in Figure 11, the variation of the droplet breakup time for the various values of flow ratios, $R$. We can clearly see that with increasing the value flow ratio ($R$), the splitting time decreases, and, any increase in the flow ratio ($R$) beyond the critical value, increases the splitting time. Thus, it may be argued that there exist a particular ferrofluid slug length at which the splitting time is minimum. For the present case, the droplet experiences a minimum splitting time for $R = 0.66$. In the inset of Figure 11, we show the variation of $W^*$ for the different values of $Bo_m$ obtained at $R = 0.66$. Although the characteristics behavior of the change in width of the droplet is almost similar, a clear distinction between the overall droplet splitting time is evident for varying magnetic field strength (as reflected by the change in $Bo_m$). The minimum splitting time can be observed for $Bo_m = 7.5$, as witnessed by the inset of Figure 11.

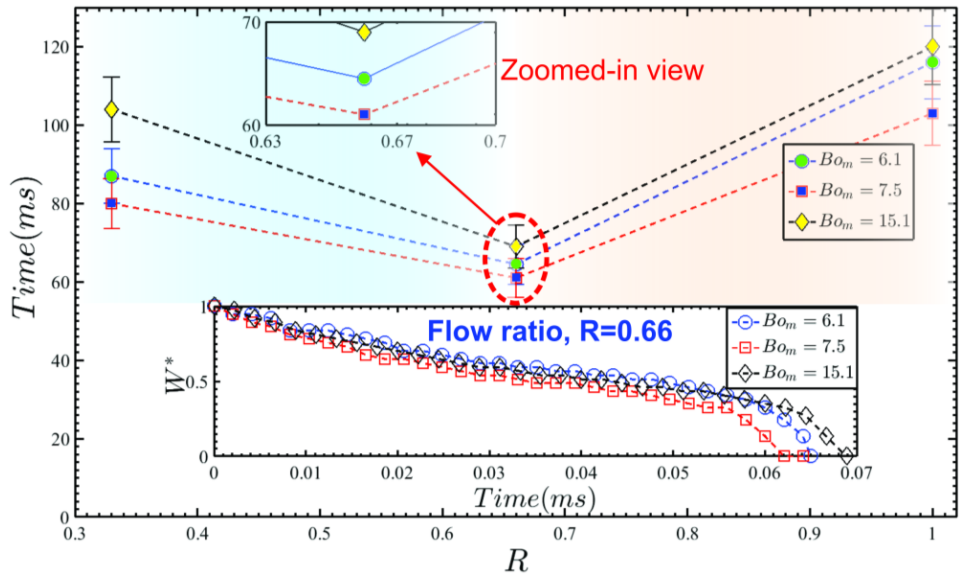

**FIGURE 11. (Color online)** Droplet splitting time ($t^*$) for the various magnetic Bond number ($Bo_m$) under consideration. The red-colored ellipse indicate the flow ratio at which the ferrofluid droplet exhibits minimum splitting time. The inset shows the variation of non-dimensionalized width ($W^*$) of the droplet for the various magnetic field intensities. $Ca = {\sim}10^{-3}$, $Re{\sim}10^{-3}$.

We show in Figure 12, the length of the sister droplet ($l_l^*$) moving in the left branch after the break up phenomena. Needless to mention here that the influence of non-uniform magnetic field is significant in the left branch. As can be clearly observed from Figure 12 that there exist a threshold magnetic field strength (precisely $Bo_m = 7.5$) and flow ratio $(R)$ at which $l_l^*$ is maximum. As can be observed from Figure 12. any increase in the magnetic field strength beyond the threshold value decreases the length of the sister droplet $(l_l^*)$ that is migrating in the left branch,. The corresponding snapshots of the events pertaining to sister droplet generation for the various cases under investigation can be observed from the inset of Figure 12. The consequence of this particular insight can be of significant importance, since by tunning the strength of the magnetic field and the flow ratio, we could ensure the desired volume of slug to move into the various branches of the T-junction divergence (precisely the left/right branch). In the suceeding section, we explore the physical reason behind this typical behaviour of the sister droplet train in a confined microfluidic passage in the presence of a non-uniform magnetic field.

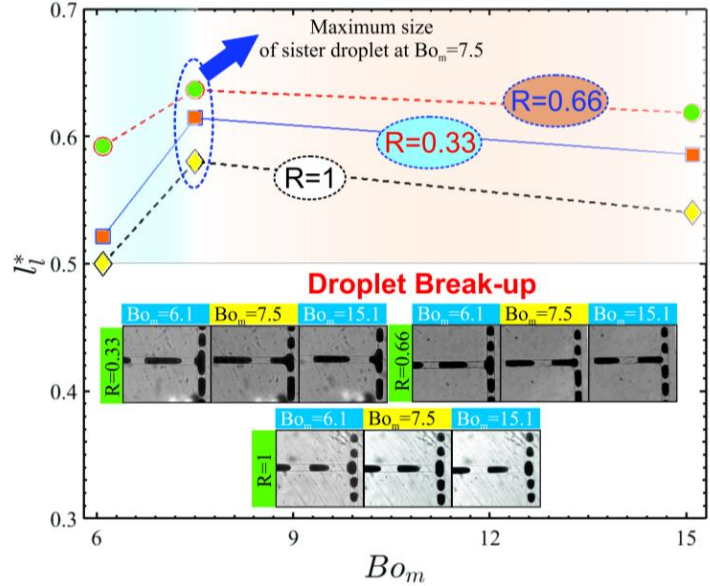

**FIGURE** 12. **(Color online)** Variation of the the non-dimensional length of the sister droplet moving in the left branch $(l_l^*)$ for the various magnetic field Bond number $(Bo_m)$. Blue colored arrow indicate the position of the threshold $Bo_m$ at which $l_l^*$ is maximum. $R = Q_c/Q_d$, where $Q_c, Q_d$ are the flow rate of the continuous phase and the dispersed phase respectively. The inset shows the snapshots of the splitted sister droplet for the various cases under consideration. $Ca = \sim 10^{-3}$, $Re \sim 10^{-3}$.

### 3.4 Mechanism of Splitting: A Time Scale Perspective

The results up till now has shown that there exists a threshold magnetic field strength $(Bo_m)$ and flow ratio $(R)$ for which the size of the generated sister droplet in the left branch becomes maximum. In this section, we attempt to unearth the reason behind this typical splitting behavior of droplets in the presence of a non-uniform magnetic field. As already mentioned before that in the present work, splitting takes place by virtue of permanent obstruction. Therefore, the underlying motion of the droplet (being splitted) moving in the left branch and the right branch simultaneously will dictate the overall splitting process. To develop an overall understanding of the splitting phenomena, we identify two parameters i.e., the velocity of the sister droplet advancing towards the left $(u_l)$, and the right divergences $(u_r)$, simultaneously. It is found that due to the non-uniformity in the distribution of magnetic field flux density, there exist a significant difference between $u_l$ and $u_r$. As a consequence of this effect, the time scale of the flow in the left branch $(t_l = l/u_l)$ and in the right branch $(t_r = l/u_r)$ of the microfluidic channel becomes different.

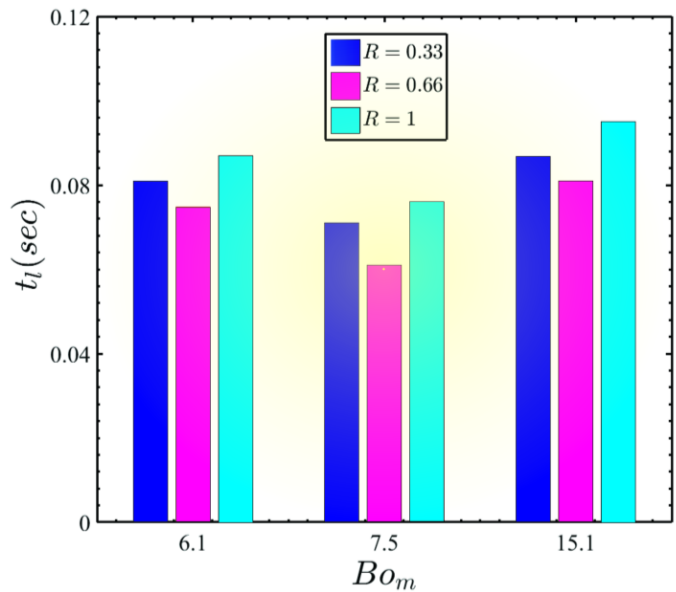

**FIGURE** 13. **(Color online)** Variation of the flow time scales of the a ferrofluid sister droplets migrating in the left branch for the various magnetic Bond number $(Bo_m)$ and flow ratios $(R)$. $t_l$ indicates the flow time scale in the left and righ branch of the T-junction respecetively.

The flow time scale refers to the time required by the sister droplet to travel the characteristics length $(l)$ in the respective branches (left/right) of the divergence i.e. the T-junction divergence. It may be reiterated that the dispersed phase (precisely the ferrofluid droplet) on impacting the T-junction divergence (i.e. the splitting junction) blocks the whole junction and stretches itself in the left and right branches respectively. During the stretching phase, the dumbbell shaped bulges of the droplet (being splitted) moves in both the left and right branches independently. This migration event eventually culminates into the breaking of the droplet into two sister droplets. However, in presence of a non-uniform magnetic field, the time scale of motion of the respective sister droplet is different. Asymmetry between the two time scales i.e. $t_l$ and $t_r$, leads to the formation of sister droplets of unequal sizes. It is understandable that due to the presence of high force field gradient in the left branch, $t_l < t_r$, indicating that the time required for the dispersed phase (i.e., the sister droplet) to move a characteristics length ($l$) is more in the right branch in comparison to the left branch. Moreover, we found that the flow time scale ($t_l$) in the left branch is dependent on the simultaneous effect of the magnetic field flux density and the flow ratio ($R$). In Figure 13, we show the variation of $t_l$, for the different cases under consideration. It can be clearly observed from Figure 13 that the migration time scale of the sister droplet moving in the left branch is minimum for $Bo_m = 7.5$ for all the cases under consideration. As a consequence, we have observed that $l_l^*$ is maximum for $Bo_m = 7.5$ (refer Figure (11)-(12) for details).

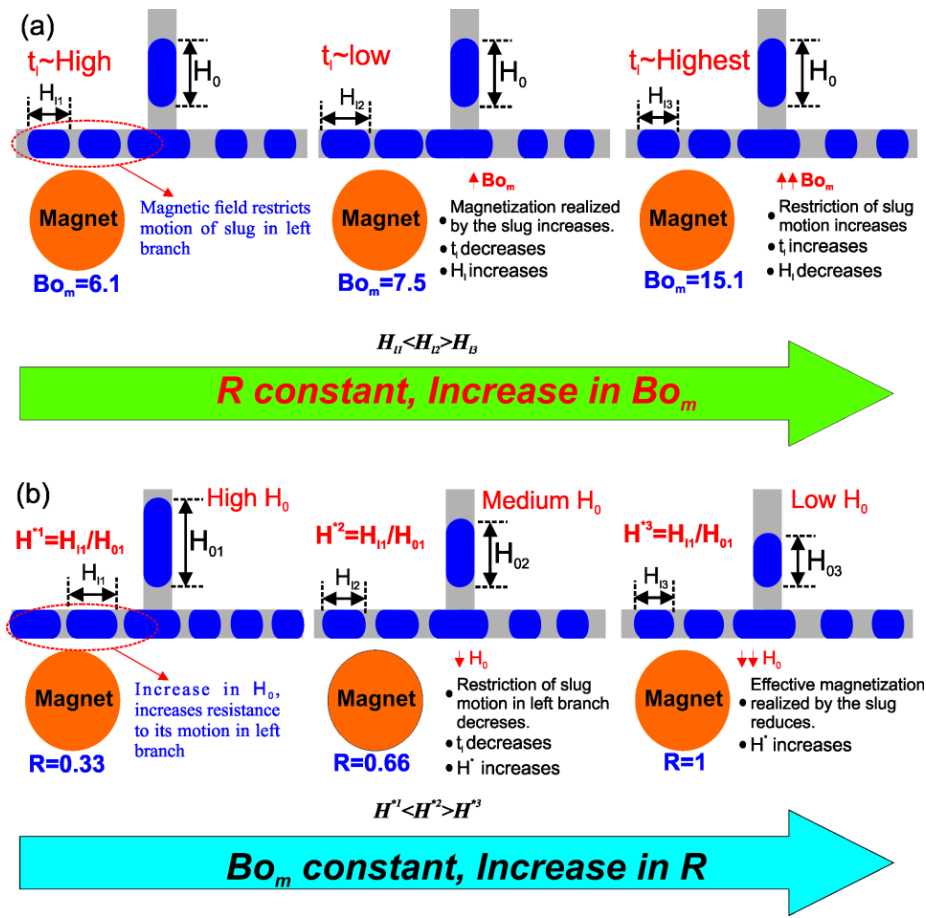

**FIGURE 14. (Color online)** Schematic representation of the droplet break up dynamics for the case when (a) $R$ is constant, $Bo_m$ increases, (b) $Bo_m$ is constant, $R$ increases.

We explain this typical behavior of the times scales in the left/right branch of the T-junction divergence with help of a graphical representations as shown in Figure 14 (a)-(b).

Although, as already discussed before, it is comprehensible that magnetic field gradient promotes on the asymmetric splitting of the ferrofluid droplets. It cannot be ignored that the flow of ferrofluid droplet train in the left branch also faces resistance due to the non-uniform distribution of the magnetic field flux density, (since the magnet is placed adjacent to the left branch). It is worth mentioning here that it is because of this resistance, the droplets moving in the left branch and the right branch has different times scale, as aptly illustrated in Figure 14(a). With an initial increase in the magnetic Bond number ($Bo_m$), the size of the ferrofluid droplets moving in the left branch ($l_f^*$) increases. With further increase in $Bo_m$ beyond the critical value, $l_f^*$ decreases. This decrease in $l_f^*$ is due to the substantial increase in the restriction to the sister droplets motion in the left branch as can be observed from Figure 14(a). This arguments gets further justified, since, we observe an increase in $t_l$ beyond critical Bond number (i.e., $Bo_{m,Crt} = 7.5$), as depicted in Figure 13. As a consequence of this resistance, the flow rate of the continuous phase in this particular branch decreases. Primarily due to this reason we observe a decrease in the size of the sister droplet (moving in the left branch) beyond the critical Bond number ($Bo_{m,Crt}$).

Similar, observation can be seen with a change in the flow ratio ($R$). Note that a low flow ratio ($R$) ensures a high initial slug length ($l_0$) of the mother droplet and vice versa. As discussed previously, an increase in the mother droplet length will ensure a simultaneous high resistance to the flow (in the left branch), thereby further decreasing $l_f^*$ as depicted in Figure 10(b). On the other hand, with a reduction in initial droplet length ($l_0$), $l_f^*$ increases. However, a further reduction in $l_0$ (i.e., increasing flow ratio, $R$) beyond the critical value ($l_{0,crt}$), limits the amount of magnetization realized by the droplet that is being splitted. This limited realization of the magnetic force by the ferrofluid droplets restricts the asymmetricities involved in the droplet splitting phenomena and leads to a decreases in $l_f^*$ further. This phenomena can be clearly observed from the graphical representation of 14(b).

## CONCLUSION

In summary, we have systematically investigated the ferrofluid droplet breakup dynamics in a T-junction divergence of LOC device in the presence of a non-uniform magnetic field. The study is especially limited to the "*breakup with permanent obstruction*" regimes. Firstly, we have methodically explored the droplet breakup behavior under the modulation of a non-uniform magnetic field. With the help of numerical simulations, we have investigated, the internal hydrodynamics of the droplet under the influence of a non-uniform forcing. We found out the presence of a "*hump-like structure*" developed inside the left moving bulge (of the ferrofluid droplet) triggers the onset of augmented flow convections inside the moving volume. By performing further investigation to this end, we have explored the role of critical magnetic Bond number $(Bo_m)$, and the flow ratio $(R)$ on the underlying droplet splitting phenomena. A critical magnetic Bond number $(Bo_{m,crt})$ is found for which the length of the splitted slug $(l_f^*)$ becomes maximum in the left branch, while any increase in magnetic field strength beyond the critical value is seen to reduce $l_f^*$. Similarly, we have also observed the presence of a threshold flow ratio $(R_{crt})$ at which the length of the splitted droplet $(l_f^*)$ is maximum. To understand these characteristics behavior of the ferrofluid droplet undergoing deformation in the presence of non-uniform force field, we have investigated the role played by the various dominant time scales. We found that due to the imbalance between the flow time scale in the left and right branches of the T-junction divergence, asymmetric droplet breakup takes place. The non-uniformity of the magnetic field (placed adjacent to the left branch) initiates a lower flow time scale in the left branch in comparison to the right branch. Also, we unveiled that at the critical state, the flow time scale is lowest in the left branch, thereby ensuring largest size of the sister droplet to be formed therein. At the critical $Bo_{m,crt}$, the length of the sister droplet (in left branch) is increased by almost 33% when compared with the base case, i.e., in absence of magnetic field. The proposed technique is expected to open-up new avenues towards the development of a LOC device for achieving rapid and controlled droplet splitting, typically finds relevance in point-of-care related diagnostics.

## ACKNOWLEDGEMENT

SS and PKM gratefully acknowledge the financial grant obtained from NEWGEN IEDC. Also, PKM gratefully to acknowledges the financial support provided by the DSIR, Govt. of India, through project no. DSIR/PRISM/170/2020-21. The authors also acknowledge the CIF, IIT Guwahati for the support in characterization of ferrofluid.